\documentclass[preprint,review,12pt]{elsarticle}

\usepackage{graphicx}
\usepackage{amssymb}
\usepackage{lineno}
\usepackage[super]{nth}

\journal{Advances in Space Research}

\begin{document}

\begin{frontmatter}


\title{Effects of CME and CIR induced geomagnetic storms on low-latitude ionization over Indian longitudes in terms of neutral dynamics}



\author[a]{Sumanjit Chakraborty\corref{c-d54cc1eb1ca4}}
\ead{phd1601121006@iiti.ac.in}\cortext[c-d54cc1eb1ca4]{Corresponding author.}
\author[c]{Sarbani Ray}
\ead{sarbanir@yahoo.com }
\author[c]{Dibyendu Sur}
\ead{dibyendumalay@gmail.com}
\author[a,b]{Abhirup Datta}
\ead{abhirup.datta@iiti.ac.in}
\author[c]{Ashik Paul}
\ead{ap.rpe@caluniv.ac.in }

\address[a]{Discipline of Astronomy, Astrophysics and Space Engineering\unskip, IIT Indore\unskip, Simrol \unskip, Indore\unskip, 453552}

\address[b]{Center for Astrophysics and Space Astronomy \unskip, Department of Astrophysical and Planetary Science \unskip, University of Colorado \unskip, Boulder \unskip, CO 80309, USA}

\address[c]{Institute of Radio Physics and Electronics\unskip,University of Calcutta\unskip, Kolkata 700 009\unskip, West Bengal, India
\\\textbf{Advances in Space Research Volume 65, Issue 1, 1 January 2020, Pages 198-213}
}

\begin{abstract}

This paper presents the response of the ionosphere during the intense geomagnetic storms of October 12-20, 2016 and May 26-31, 2017 which occurred during the declining phase of the solar cycle 24. Total Electron Content (TEC) from GPS measured at Indore, Calcutta and Siliguri having geomagnetic dips varying from 32.23$^\circ$N, 32$^\circ$N and 39.49$^\circ$N respectively and at the International GNSS Service (IGS) stations at Lucknow (beyond anomaly crest), Hyderabad (between geomagnetic equator and northern crest of EIA) and Bangalore (near magnetic equator) in the Indian longitude zone have been used for the storms. Prominent peaks in diurnal maximum in excess of 20-45 TECU over the quiet time values were observed during the October 2016 storm at Lucknow, Indore, Hyderabad, Bangalore and 10-20 TECU for the May 2017 storm at Siliguri, Indore, Calcutta and Hyderabad. The GUVI images onboard TIMED spacecraft that measures the thermospheric O/N$_2$ ratio, showed high values (O/N$_2$ ratio of about 0.7) on October 16 when positive storm effects were observed compared to the other days during the storm period. The observed features have been explained in terms of the O/N$_2$ ratio increase in the equatorial thermosphere, CIR-induced High Speed Solar Wind (HSSW) event for the October 2016 storm. The TEC enhancement has also been explained in terms of the Auroral Electrojet (AE), neutral wind values obtained from the Horizontal Wind Model (HWM14) and equatorial electrojet strength from magnetometer data for both October 2016 and May 2017 storms. These results are one of the first to be reported from the Indian longitude sector on influence of CME- and CIR-driven geomagnetic storms on TEC during the declining phase of solar cycle 24. 
\end{abstract}


\end{frontmatter}

% \linenumbers

\section{Introduction}

Ionospheric storms describe variations in the ionosphere due to geomagnetic disturbances. These storms occur due to the input of solar wind energy, which is sudden, into the magnetosphere-ionosphere-thermosphere system. Generally, a geomagnetic storm occurs following a Coronal Mass Ejection (CME) when the polarization of Interplanetary Magnetic Field (IMF) Bz changes from northward to southward, remain southward for several hours \citep{sc:21} and reconnects with the Earth's magnetic field \citep{sc:14}. During the declining phase of the solar activity, another important phenomenon that causes geomagnetic storms is the coronal holes. Coronal holes emanate high speed solar wind ($>$750-800 km/s) streams which overtakes and interacts with slow ($\sim$300-400 km/s) streams near the ecliptic plane to produce regions of intense magnetic field called Corotating Interaction Region (CIR) \citep{sc:43}. The magnetic storms caused by CIRs are of weak to moderate intensity because the z component of the Geocentric Solar Magnetospheric (GSM) magnetic field within the CIR is of oscillatory nature \citep{sc:42}. The southward negative component of Alfv\'en waves contained in the high speed streams reconnects with Earth's magnetic field to transfer energy from the solar wind to the magnetosphere-ionosphere-thermosphere system \citep{sc:10,sc:41,sc:22,sc:44}. 

CME-related magnetic storms refer to the storms driven by components such as the interplanetary shocks, the strong magnetic field of the sheath and the magnetic field in the ejecta. Since the HSSW streams follow the CIRs, so either or both are drivers of the CIR-related storms. CME-driven storms are dominant at solar maximum while the CIR-driven storms tend to occur during the declining phase of a solar cycle. Occurrence pattern of the CME-driven storms is irregular, with highest occurrence frequency during solar maximum whereas CIR-driven storms repeat with a period of about 27 days as a result of Sun's rotation, mainly during a solar cycle's declining phase. Furthermore, CME-driven storms have a shorter recovery phase lasting for about 1–2 days while the CIR-driven storms have a prolonged recovery phase that can last for several days. Storm Sudden Commencements (SSC) are associated with the interplanetary shocks driven by the CMEs while the gradual onsets are driven by HSSW streams \citep{sc:46}. 

The beginning of geomagnetic disturbances is associated with large variations of the IMF Bz and Prompt Penetration Electric Fields (PPEF) which results from the transmission of magnetospheric convection electric field from high to low latitudes \citep{sc:45}. The equivalent current system which is associated with the prompt penetration \citep{sc:30} has average periods of the order of few minutes to 3 hours. During the previous decades, both theoretical and experimental studies have clearly highlighted the characteristics of prompt penetration of magnetospheric convection electric field from high to low latitudes and the severe effects on low latitude electrodynamics \citep{sc:15,sc:40,sc:16, sc:9}.

At the higher latitudes, Auroral Electrojet (AE) currents transfer energy in the form of heat, by the physical process of Joule heating, to the neutral gas. This Joule heating is given by the power per unit volume dP/dV = J.E, (where J is the electric current and E the electric field). Additionally, by the process of Ampere force given by cross product of J and B, where B is the magnetic field, AE currents move the neutral wind by transfer of momentum.  Electric fields and currents get generated by the changes in the neutral thermosphere circulation, at the origin of the ionospheric disturbance dynamo \citep{sc:39,sc:20}. The ionospheric disturbance dynamo is a long lasting process (from hours to days). Unlike PPEF which is eastward during dawn-dusk, the ionospheric Disturbance Dynamo Electric Field (DDEF) is eastward in night side while westward in dayside. These effects vary slowly and are present for a long duration of time even after the subsiding of magnetic activity as a consequence of the inertia from the Coriolis force \citep{sc:7}. 

The low latitude ionosphere is characterized by Equatorial Ionization Anomaly (EIA). The anomaly starts at 09:00 LT and by 16:00 LT moves poleward up to 15$^\circ$ dip latitude and finally comes back towards the dip equator and vanishes by 24:00 LT \citep{sc:34,sc:2,sc:36}. The EIA is caused by two main plasma motions at the F region. Firstly, the motion that is perpendicular to the magnetic field of the Earth, generates plasma drift directed upwards caused by the zonal eastward(day-time)/westward(night-time) electric field produced by the E region dynamo and B. The second motion which is parallel to B, as a result, under the influence of ambipolar diffusion drift related to pressure and gravity gradients, causes plasma to diffuse following the geomagnetic field lines \citep{sc:24}. The day-to-day variability and dynamics of the EIA and consequently the Total Electron Content (TEC) are well known to be largely due to thermospheric winds and electric fields \citep{sc:3}. The storm time variability of EIA has been addressed by several workers \citep{sc:30,sc:31,sc:11,sc:23,sc:5,sc:38}.

The enhanced eastward PPEF raises the dayside and evening ionosphere upward significantly causing huge increase in the ionospheric TEC \citep{sc:27}. During such periods, there is a poleward extension of the EIA with its crest directing into the mid-latitudes \citep{sc:4}. During geomagnetic storms, the thermosphere and ionospheric structures are affected by ion drag forcing and Joule heating \citep{sc:6}. This energy input at high latitudes modifies the ionospheric F region electron density. Hence the electric field changes globally as a result of these disturbances and produces an increase in the electron density which is well known as the positive storm effect \citep{sc:29}.

Considerable amount of work on ionospheric TEC response to geomagnetic storms have been reported in literature. In recent years, \citep{sc:6} studied the ionospheric response to the St. Patrick's Day storm of March 16-21, 2015, which is the most intense (G4 class, Kp=8, severe, according to NOAA scales http://swpc/noaa.gov/noaa-scales-explanation) storm of solar cycle 24. They have observed dramatic TEC enhancements in and around the area of the eastern Pacific region. For the same storm, \citep{sc:28} using TEC measurements from the middle and low latitudes in the Pacific, American, African and Asian longitude sectors showed positive storm effects during the main phase of the storm depending on the longitude sector. During the recovery phase, sharp decrease in ionization lasting for several days has been observed at all longitude sectors. \citep{sc:35} have reported observations of amplitude, phase scintillations and TEC from different GPS stations of India during the storm of 17 March, 2015 and highlighted its effects on the performance of GPS. A very interesting observation of severely depleted ionosphere on the day following the St.Patrick day storm of 2015 at high latitudes (north of 65$^\circ$N) was reported by \citep{sc:32} using GPS TEC measurements over a network of stations extending from the equator to the polar regions along the African and European longitude sector. The Indian subcontinent covers the low latitude zone in the South-Asian longitudes where the magnetic equator passes over the southern tip of the peninsula near Trivandrum \citep{sc:13}.

The paper presents the effect of the October 2016 storms that was both CME and CIR-induced and May 2017 storm that was CME-induced over the Indian longitude sector. The stations have been carefully selected to reveal changes in the dynamism of the ionosphere in such an extremely variable ionospheric region during geomagnetic disturbances. The paper also shows the effect of HSSW on the ionospheric TEC at different locations in and around the EIA in the Indian sector. The novelty of this work lies in the fact that impact of combined CME- and CIR-driven geomagnetic storms on ionospheric Global Navigation Satellite System (GNSS) TEC over the Indian subcontinent has not been extensively reported in literatures. Furthermore, the locations of the different observing stations ensure characterization of storm-induced effects on ionospheric TEC near the magnetic equator (Bangalore) as well as around the northern crest of the EIA (Indore and Calcutta) and beyond (Lucknow and Siliguri).

\section{Data and Methods}

\begin{figure}
\centering\includegraphics[width=1\linewidth]{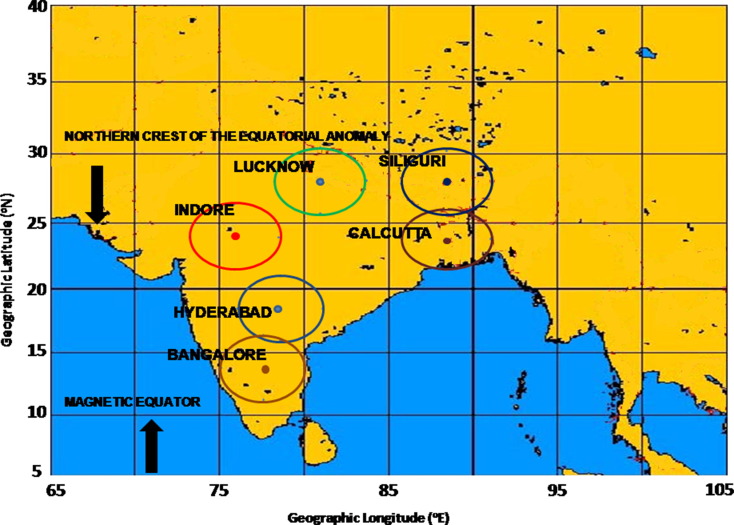}
\caption{Locations of the stations Bangalore, Hyderabad, Indore, Calcutta, Siliguri and Lucknow on a map of India. Zone of reception from these stations above an elevation of 50° is shown by the different colored ellipses. The magnetic equator and the location of the northern crest of the equatorial ionization anomaly (EIA) are indicated in the map.}
\label{sc1}
\end{figure}

A multi-constellation and multi-frequency Global Navigation Satellite System (GNSS) receiver is operational since May 2016 at the Discipline of Astronomy, Astrophysics and Space Engineering, Indian Institute of Technology Indore, Indore (22.52$^\circ$N, 75.92$^\circ$E geographic; magnetic dip 32.23$^\circ$N) which is located in the EIA region of the Indian longitude sector. This GNSS receiver has the capability of tracking GPS, GLONASS, GALILEO and SBAS geostationary satellites at multiple frequencies (L1 – 1575.42MHz, L2 – 1227.6MHz, L5 – 1176.45MHz). The output includes the azimuth and elevation of the satellite, time (UTC) and calibrated TEC recorded at 1 minute sampling interval. Data has also been used from similar receivers operational at the Institute of Radio Physics and Electronics, University of Calcutta, Calcutta (22.58$^\circ$N, 88.38$^\circ$E geographic; magnetic dip 32$^\circ$N) and North Bengal University, Siliguri (26.72$^\circ$N, 88.39$^\circ$E geographic; magnetic dip 39.49$^\circ$N). Ionospheric TEC data, available at a sampling interval of 1 min from the IGS stations at Lucknow (26.91$^\circ$N, 80.95$^\circ$E geographic; magnetic dip 39.75$^\circ$N), Hyderabad (17.41$^\circ$N, 78.55$^\circ$E geographic; magnetic dip 21.69$^\circ$N) and Bangalore (13.02$^\circ$N, 77.5$^\circ$E geographic; magnetic dip 11.78$^\circ$N), have also been analyzed available at (http://sopac.ucsd.edu/). Figure \ref{sc1} shows the zone of reception above an elevation angle of 50$^\circ$ for all the stations presented in this paper along with the northern crest of EIA and the magnetic equator that passes over Tirunelveli (8.73$^\circ$N, 77.70$^\circ$E geographic; magnetic dip: 3.26$^\circ$N). The present study emphasizes on TEC measurements obtained from a distribution of stations located at Lucknow, beyond the northern crest of EIA, Indore (located near the crest of EIA), Hyderabad (located between geomagnetic equator and northern crest of EIA) and Bangalore (located near magnetic equator) for the October 2016 storm. The stations used for the May 2017 storm are Siliguri (located beyond the crest of EIA), Calcutta (located near the EIA crest), Indore and Hyderabad.

In the analyses, Ionospheric Pierce Point (IPP) height of 350 km has been used. The hourly Disturbance storm time (Dst) (nT) index has been obtained from World Data Center for Geomagnetism, Kyoto (http://wdc.kugi.kyoto-u.ac.jp/dstdir/). The storm indices Auroral Electrojet AE (nT), IMF Bz (nT), the Solar Wind Speed (km/s) and Interplanetary Electric Field Ey (mV/m), are obtained from NASA's Space Physics Data Facility (SPDF) omniweb service (http://omniweb.gsfc.nasa.gov) at 1 minute resolution and have been plotted for the disturbed period of October 12-20, 2016 and May 26-31, 2017. 

Since 2002, GUVI on-board the TIMED spacecraft provides the thermospheric column number density ratio of O and N$_2$  at N$_2$ column number density level of 10$^{17}$ cm$^{-2}$ (O/N$_2$ ratio) \citep{sc:26}. The GUVI O/N$_2$ ratio has been derived from the measurements of OI 135.6 nm and N$_2$ Lyman-Birge-Hopfield airglow emissions. The O/N$_2$ ratio is widely used to describe the effect of thermospheric neutral composition change on the ionosphere. The global variation of O/N$_2$ ratio during the storm periods, have been taken from http://guvitimed.jhuapl.edu/.

The meridional and zonal neutral wind velocities described in this paper are obtained from the Horizontal Wind Model (HWM14). The model provides temporal and spatial profiles of upper atmospheric winds and tides. It is a data assimilated empirical model consisting of large dataset from Fabry-Perot interferometer, satellite measurements, sounding rockets, medium-frequency radar, incoherent backscatter radar and also from other wind-prediction systems \citep{sc:47,sc:48}. In order to find the disturbed components of neutral wind, the model uses a subroutine called Disturbance Wind Model (DWM) \citep{sc:49}. Furthermore, to estimate the strength of equatorial electrojet, ground magnetic data are obtained from the INTERMAGNET available at http://www.intermagnet.org/index-eng.php.

\section{Results and Discussions}

\subsection{Storm of October 12-20, 2016}

A magnetic filament in the Sun's northern hemisphere erupted on October 08, 2016, around 16:00 UT hurling a CME into space. The CME with a weak shock arrived at 21:21 UT on October 12, 2016 as observed by Deep Space Climate Observatory (DSCOVR) at L1-point (http://www.spaceweather.com). As a result, a G2 class (Kp = 6, moderate, according to NOAA scales http://swpc/noaa.gov/noaa-scales-explanation) geomagnetic storm commenced at 08:15 UT on October 13, 2016. On October 13, NASA's Solar Dynamics Observatory (SDO) captured dark coronal hole which almost pointed directly at the Earth. At the leading edge of the emerging solar wind stream, a CIR was observed as the Earth entered a fast moving solar wind stream on October 16, 2016 (http://www.spaceweather.com). 

\begin{figure} 
\centering\includegraphics[width=0.75\linewidth]{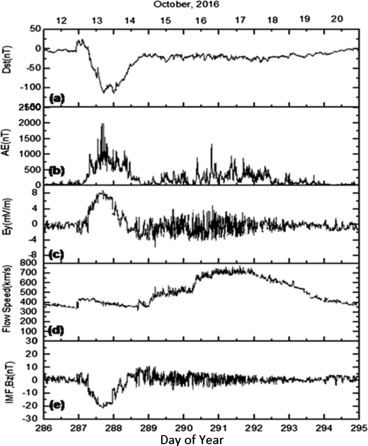}
\caption{Variation of the Dst index, Auroral Electrojet (AE), Interplanetary Electric Field (Ey), Solar Wind Speed (V) and IMF Bz during October 12-20, 2016.}
\label{sc2}
\end{figure}

Figure \ref{sc2} shows the different geomagnetic storm indices for the storm period of October 12-20, 2016. The minimum value of Dst (Figure \ref{sc2}a) reached -103 nT at 18:00 UT on October 13, 2016. Figure \ref{sc2}b shows the AE index which is the difference between the strongest current intensity of the eastward and westward auroral electrojet. AE index first peaked at 17:00 UT on October 13, 2016 with a value of 1992 nT and had a second peak at 19:00 UT on October 16, 2016 with a value of 1333 nT. The first peak of AE that was observed on October 13, 2016 is due to the CME-induced storm while the second peak which was observed on October 16, 2016 is due to the CIR related magnetic storm \citep{sc:50}. The AE index remained above 700 nT for about an hour during 13:00 UT to 17:00 UT on October 13, 2016. In Figure \ref{sc2}d, it is observed that the average flow speed was about 367 km/s, 411 km/s and 369 km/s on October 12, 13 and 14 respectively. On October 15, the solar wind speed scaled up in steps from 384 km/s at 0:00 UT to 530 km/s at 23:00 UT. On October 16, a sudden jump in solar wind speed from a value of 571 km/s at 07:00 UT to 642 km/s at 08:00 UT was observed as the Earth entered a HSSW stream. The flow speed reached a peak of 777 km/s at 14:21 UT on October 17 but its average value was 628 km/s on this day. October 18 and 19 noted average flow speed values of 630 km/s and 501 km/s respectively. The speed gradually went down to an average value of 401 km/s on October 20, 2016. Figure \ref{sc2}e shows IMF Bz component which remained below -10 nT for a period starting from 08:26 UT to 21:58 UT on October 13, 2016 and reached a peak $\sim$21nT at 15:18 UT, October 13. The IMF Bz started oscillating from October 14 till October 17 further recording a value of -7.32 nT on October 16 at 06:00 UT. The derived Interplanetary Electric Field (Ey) has been plotted in Figure \ref{sc2}c. The peak Ey of value 8.77 mV/m was recorded on October 13 at 16:18 UT during the main phase of the storm corresponding to the southward turning of IMF Bz on the same day. 

\begin{figure} 
\centering\includegraphics[width=1\linewidth]{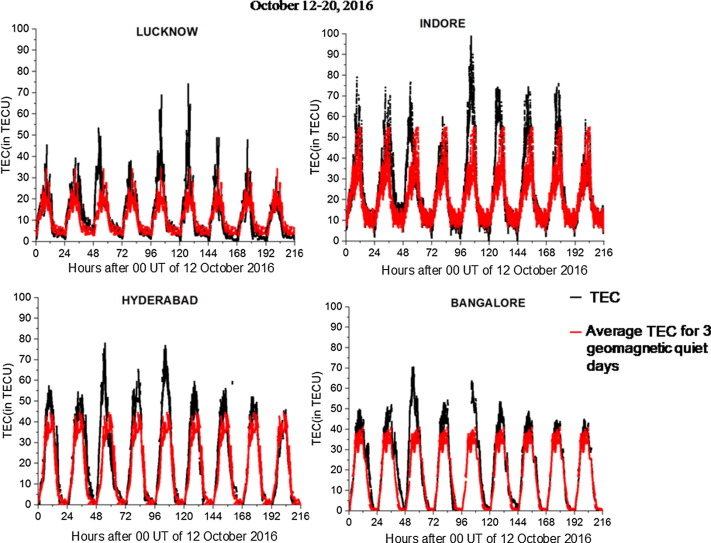}
\caption{TEC diurnal variation recorded at Lucknow, Indore, Hyderabad and Bangalore as a function of Universal Time (UT) along with the average TEC for three geomagnetic quiet days.}
\label{sc3}
\end{figure}

In order to understand the effects of geomagnetic storms over a distributed network of GPS stations in the Indian longitude sector on TEC, diurnal variations of TEC have been studied from different stations. Figure \ref{sc3} shows the diurnal plots of TEC that includes all satellites with elevation angle greater than 50$^\circ$ for the stations Lucknow, Indore, Hyderabad and Bangalore during October 12-20, 2016 plotted with reference to the average TEC for geomagnetic quiet condition. The cut-off for elevation angle has been chosen to be 50$^\circ$ so as to minimize the effects of sharp spatial gradients of ionization which occur in the equatorial and low latitudes.  The mean for three geomagnetic quietest days of the month of October viz. \nth{11}, \nth{20} and \nth{21}, chosen based on the lowest Sigma Kp values of the month (http://wdc.kugi.kyoto-u.ac.jp/kp/index.html), are plotted in red in this figure. During the recovery phase of the storm i.e. during October 14-20, prominent peaks in diurnal maximum in excess of 20-45 TECU over the quiet time values (that would introduce range errors of 3.2-7.2 m on GPS L1 frequency) were observed from Lucknow on October 14, 16 and 17, from Indore on October 14 and 16, from Hyderabad on October 14, 15, and 16 and Bangalore on October 14 and 16 signifying a positive storm effect. This significant enhancement in TEC due to the CIR-induced storm agrees with the results published by \citep{sc:51} where they report that even CIR-induced magnetic storms of weak to moderate intensity can have effects on the ionosphere in a way similar to the effects caused due to the strong storms.

Joule and particle heating of the polar upper atmosphere causes upwelling of the molecular species N$_2$ and O$_2$ and then pressure gradient force and ionospheric convection extend the molecule rich gases equatorwards \citep{sc:25}. The global wind circulation induced by heating of the polar upper atmosphere during storm period causes downwelling of the molecular gases in the mid- and low latitudes, thereby reducing the O+ loss rate in the F region in these latitudes and causing positive storm effect \citep{sc:18,sc:19,sc:12,sc:17}.

\begin{figure}
\centering\includegraphics[width=1\linewidth]{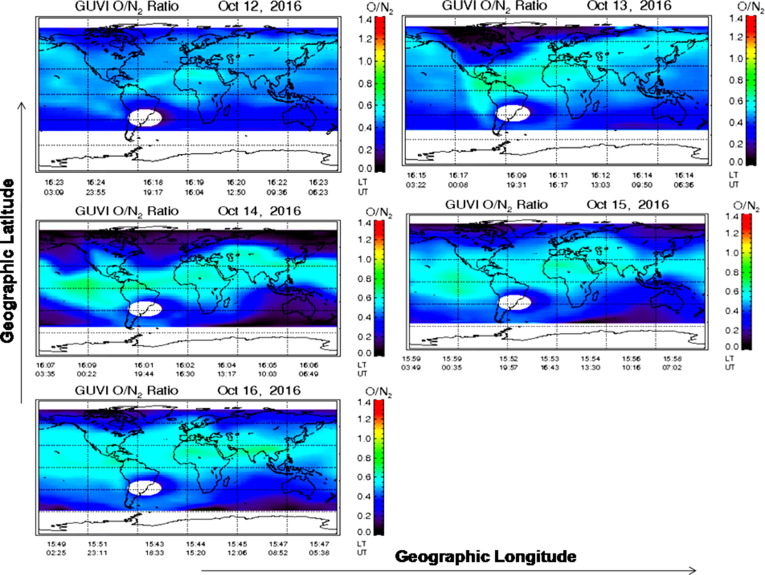}
\caption{World map showing GUVI O/N$_2$ ratio for the period October 12-16, 2016.}
\label{sc4}
\end{figure}

\begin{figure}
\centering\includegraphics[width=1\linewidth]{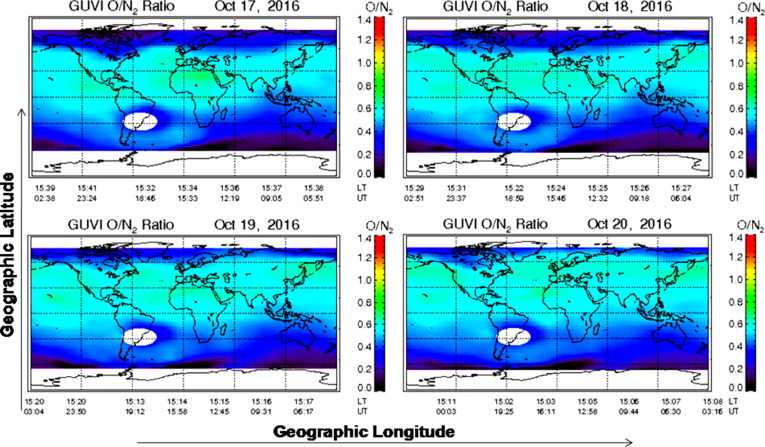}
\caption{World map showing GUVI O/N$_2$ ratio for the period October 17-20, 2016.}
\label{sc5}
\end{figure}

\begin{figure}
\centering\includegraphics[width=1\linewidth]{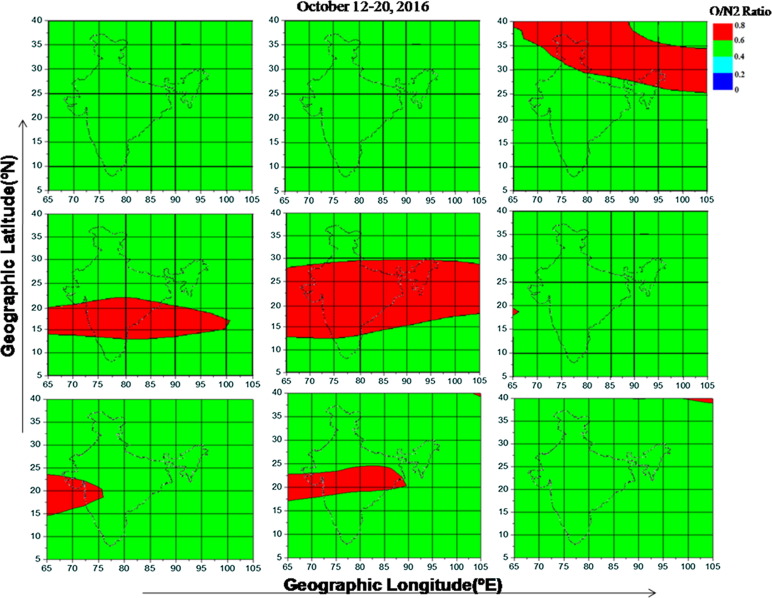}
\caption{Variation of O/N$_2$ ratio over India for the period October 12-20, 2016.}
\label{sc6}
\end{figure}

Figures \ref{sc4}, \ref{sc5} show the global variation of thermospheric O/N$_2$ ratio for the storm periods of October 12-16 and October 17-20, 2016 respectively at a nearly constant local time of 16:00. It is observed that, over the Indian longitude sector, the maximum thermospheric O/N$_2$ ratio is observed on October 16. Plotting the Indian longitude sector separately in Figure \ref{sc6}, it is observed that O/N$_2$ ratio values greater than 0.6 have been observed on October 14, 15 and 16. The maximum value ($\sim$0.7) of O/N$_2$ ratio has been observed on October 16. On October 17, the O/N$_2$ values were below 0.6 over the whole subcontinent. On October 18 and 19, there were values of O/N$_2$ greater than 0.6. October 20 was quiet without any O/N$_2$ increase above 0.6.   

The positive storm effect reported in the present paper observed in response to the CIR event of October 2016 storm from stations in the Indian longitude sector show TEC enhancement with increase in O/N$_2$ ratio on October 16. The peak of AE ($\sim$1333 nT) is observed at 19:00 UT on October 16. These also signify enhanced Joule heating on that day as the Joule heat production rate is linearly related to AE \citep{sc:1}. Therefore, duration of high AE in the form of DDEF could be an explanation to the TEC enhancements observed on some of the days of this storm. 

Neutral wind plays a pivotal role on the total ionization in low latitudes. Equatorward neutral wind reduces ion-recombination and increases the total O/N$_2$ ratio \citep{sc:52,sc:53}. An enhancement of TEC near the equatorial region decreases the sharp latitudinal TEC gradient in the low latitudes. This causes negative ionospheric storm near EIA crests. The phenomenon simultaneously causes positive ionospheric storm near geomagnetic equator \citep{sc:57,sc:56}. Neutral wind contributes towards the equatorial E$\times$B plasma drift. Any such influence on vertical plasma drift may cause change in ion-recombination rate and change in overall TEC \citep{sc:54,sc:55}. In the paper, the effects of neutral wind on TEC are observed in geomagnetic disturbed conditions in low latitude region.

\begin{figure}
\centering\includegraphics[width=1\linewidth]{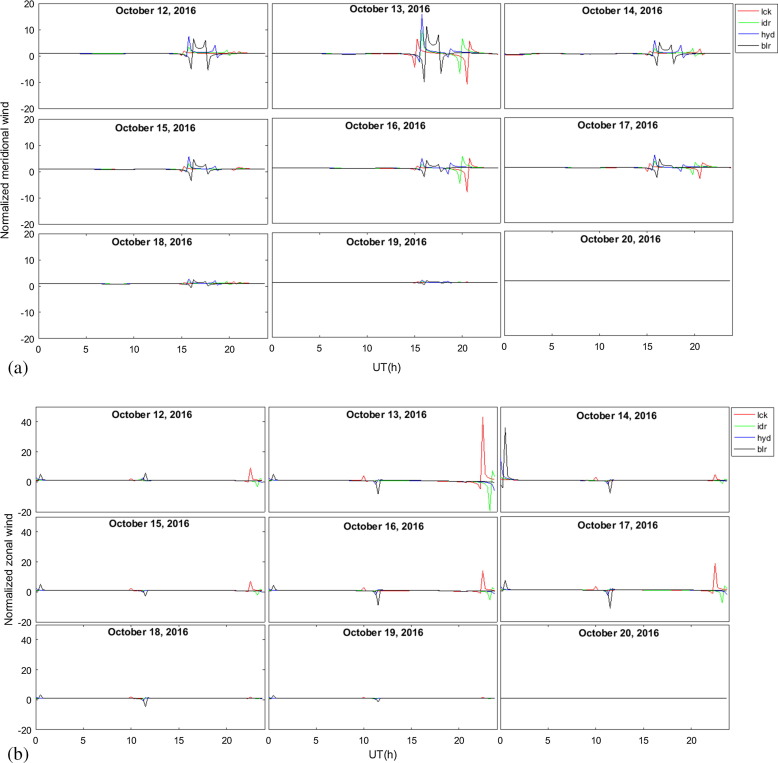}
\caption{Normalized diurnal variations of (a) meridional neutral wind and (b) zonal neutral wind during October 12–20, 2016 for the stations Lucknow (lck in red), Indore (idr in green), Hyderabad (hyd in blue) and Bangalore (blr in black). (For interpretation of the references to color in this figure legend, the reader is referred to the web version of this article.)}
\label{sc7}
\end{figure}

Figure \ref{sc7}a and b respectively shows the normalized meridional and zonal neutral wind plots, w.r.t. the quietest day (October 20, 2016) in this period. From Figure \ref{sc7}a, it can be observed that the meridional wind was negative at Lucknow during 20-21 UT on October 13 and 16, 2016 that signifies southward directed wind which bring ionized particles from high latitudes to the Indian low latitudes suggesting a high value of TEC at Lucknow on October 14 and 17, 2016. From Figure \ref{sc7}b, on October 13 and 16, 2016 around 21–22 UT, high values of zonal wind were observed from Lucknow, which signifies enhancement of eastward zonal wind in early morning (LT = UT + 5.5) of October 14 and 17, 2016. This eastward zonal wind shifts the ions eastward which enhances eastward equatorial electrojet thus enhancing the net upward E$\times$B drift. As a result, the upward movement of ions, signifying lesser recombination, contribute towards enhancement of TEC on October 14 and 17, 2016 at Lucknow. From Bangalore, sharp enhancement of eastward zonal wind is observed during 0-2 UT (LT =UT + 5) and a moderate enhancement is observed from Hyderabad justifying a TEC enhancement for these two stations on October 14, 2016. From Indore, on October 13, 2016, at 21–23 UT, a negative enhancement of zonal wind is observed which signifies the zonal wind that bring ions to westward direction at an altitude of 350 km. Thus it adds component to westward equatorial electrojet at nighttime and the net downward E$\times$B drift is strengthened. Hence at Indore no additional enhancement in TEC is observed on October 14, 2016 compared to the previous days.

\begin{figure}
\centering\includegraphics[width=1\linewidth]{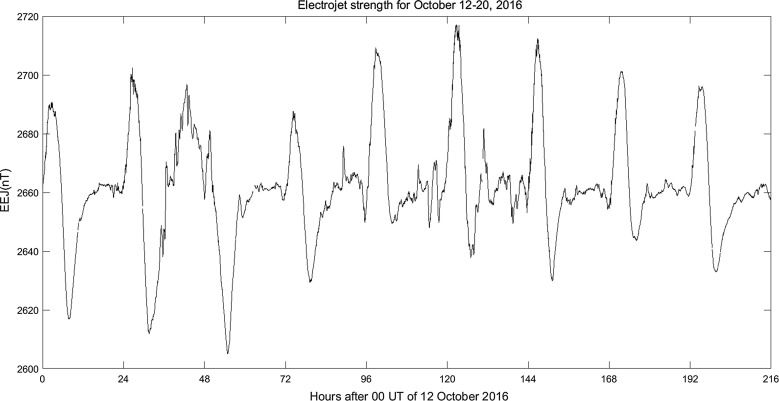}
\caption{Electrojet strength during October 12–20, 2016, taking the difference of horizontal component of the Earth's magnetic field (H in nT) of Alibag (station away from magnetic equator) from that of Dalat (a station near magnetic equator).}
\label{sc8}
\end{figure}

The equatorial electrojet strength that is related to the electric field, has been computed by taking difference of the horizontal component, H (nT) of Earth's magnetic field at Alibag (10.51$^\circ$N geomagnetic, a station away from magnetic equator), from that at Dalat (2.23$^\circ$N geomagnetic, a station near to the magnetic equator). From Figure \ref{sc8}, it is observed that this difference increased from a value $\sim$2687 nT on October 15 to values $\sim$2709 nT and 2717 nT on October 16 and 17 respectively, signifying increase of the E$\times$B force on the equatorial ionization and subsequent ionization distribution to the off-equatorial regions. This is supported by enhanced TEC over the stations as observed in Figure \ref{sc3}.

Another important feature of the October 2016 storm was the presence of CIR induced storm on October 16, in addition to the CME induced storm on October 13-14. The HSSW event observed on October 16 is characterized by an abrupt rise of solar wind speed above 700 km/s and oscillatory nature of IMF Bz with a negative peak resulting in a peak Ey. This strengthens the daytime eastward zonal electric field and causes an increase in TEC \citep{sc:37}. The enhancement in TEC continued till October 17 as long as the solar wind maintained high values above 700 km/s. Thus enhancement in TEC observed on October 16 was a consequence of HSSW arising out of coronal holes and CIR in addition to the thermospheric composition change as explained earlier. For October 17, the enhancement was due to HSSW only as there were no significant changes in O/N$_2$ on that day.
 	
It should also be noted that the positive storm effect observed for the CME-induced storm on October 13–14 was less significant than that observed during the CIR-induced storm on October 16–17. This is due to the fact that the onset of the storm at 23:00 UT on October 12 was not in the local (LT = UT + 05:30) morning to noon sector in which the ionization production mechanism dominates and plasma accumulation due to mechanical effect of neutral wind is greater than its chemical loss due to recombination \citep{sc:33,sc:8,sc:35}. 

\clearpage
\subsection{Storm of May 26-31, 2017}

\begin{figure}
\centering\includegraphics[width=0.75\linewidth]{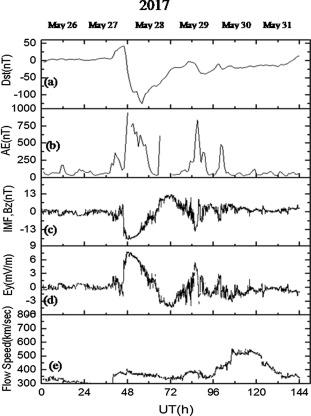}
\caption{Variation of the Dst index, Auroral Electrojet (AE), IMF Bz, Interplanetary Electric Field (Ey) and Solar Wind Speed (V) during May 26-31, 2017.}
\label{sc9}
\end{figure}

A G3 (Strong Kp=7) geomagnetic storm started at 04:19 UT due to the activity associated with the CME of May 23, 2017 (http://www.swpc.noaa.gov/
news/g3-strong-geomagnetic-storm-alert-issued). Figure \ref{sc9} shows the different geomagnetic storm indices for the storm days of May 26-31, 2017. The storm commenced at 16:00 UT on May 27. Dst (Figure \ref{sc9}a) reached a minimum value of -125 nT at 08:00 UT on May 28, 2017 and returned to a quiet value of -5 nT at 12:00 UT on May 29, 2017. A mild storm commenced at 10:00 UT on May 29, reached a minimum of -39nT at 19:00 UT and was being restored to quiet conditions on May 31 at 16:00 UT. Figure \ref{sc9}b shows the AE index which first peaked at 00:00 UT on May 28, 2016 with a value of 949 nT and had a second peak at 15:00 UT on May 29, 2017 with a value of 839 nT. Figure \ref{sc9}c shows IMF Bz component which remained below -10 nT for a period starting from 21:46 UT of May 27 to 07:34 UT of May 28 reaching a peak $\sim$ -20 nT at 00:14 UT on May 28. Then again from 13:16 - 14:24 UT on May 29, IMF Bz peaked to $\sim$ -14 nT at 13:33 UT. Figure \ref{sc9}d shows the corresponding Ey which maximized to 7.86 mV/m at 00:14 UT on May 28 and 5.51 mV/m at 13:33 UT on May 29. Figure \ref{sc9}e plots the solar wind for the storm period. The speed remained below 500 km/s on most of the days except May 30 when it peaked to 554 km/s.

\begin{figure}
\centering\includegraphics[width=1\linewidth]{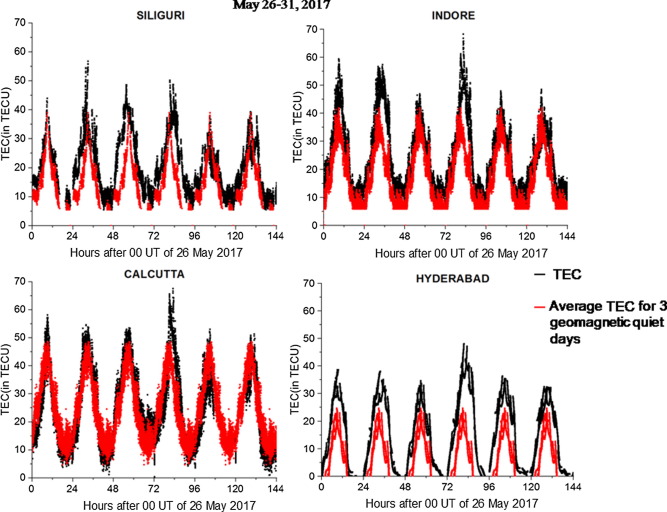}
\caption{Diurnal variation of TEC recorded at Siliguri, Indore, Calcutta and Hyderabad as a function of Universal Time (UT) along with the average TEC for three geomagnetic quiet days for the period May 26-31, 2017.}
\label{sc10}
\end{figure}

Figure \ref{sc10} shows the diurnal plots for the stations Siliguri, Indore, Calcutta, and Hyderabad during May 26-31, 2017 with respect to mean TEC for geomagnetic quiet condition, plotted in red; the three geomagnetic quietest days of the month being \nth{3}, \nth{25} and \nth{26}. Here also cut-off for elevation angle has been chosen to be 50$^\circ$. On May 27 and May 29, prominent peaks in diurnal maximum in excess of 10-20 TECU over quiet time values (that would introduce a range error of 1.6-3.2m at GPS L1 frequency), were observed from all the stations. Since the storm commenced at 16:00 UT on May 27, after the diurnal maximum, the only enhancement in diurnal TEC maximum that occurred during the storm was on May 29. On May 28, a decrease in the diurnal maximum was observed from the pre-storm days of May 26 and 27.

\begin{figure}
\centering\includegraphics[width=1\linewidth]{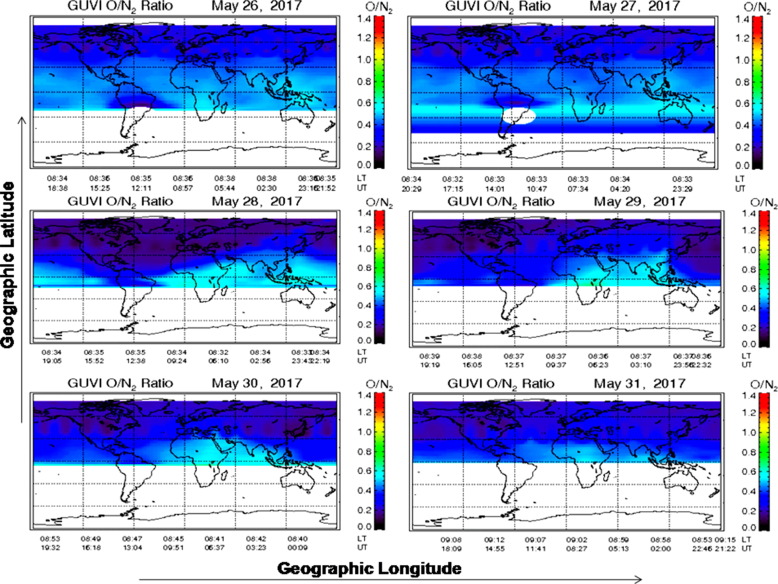}
\caption{World map showing GUVI O/N2 ratio for the period May 26-31, 2017.}
\label{sc11}
\end{figure}

\begin{figure}
\centering\includegraphics[width=1\linewidth]{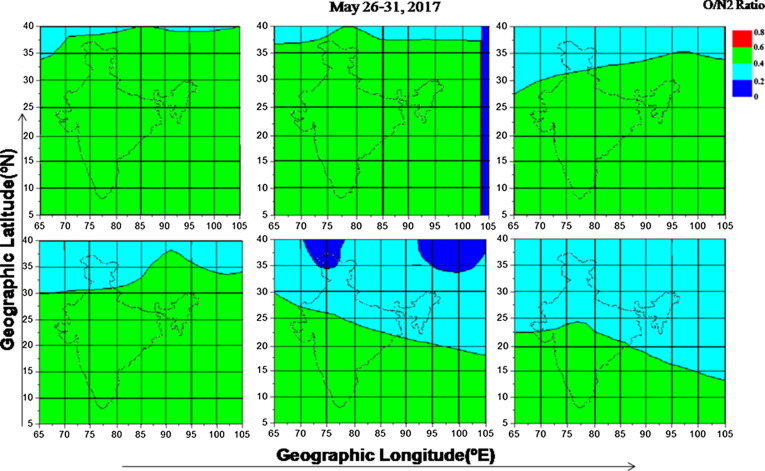}
\caption{Variation of O/N2 ratio over India for the period May 26-31, 2017.}
\label{sc12}
\end{figure}

Figure \ref{sc11}, which is similar to Figures \ref{sc4} and \ref{sc5} but for the storm of May 26-31, 2017, shows the GUVI images for the storm days to show how the thermospheric O/N$_2$ ratio varied globally at a nearly constant local time of 16:00. Over the Indian longitude sector, the thermospheric O/N$_2$ ratio is nearly similar for all the days. Concentrating on the Indian longitude sector (Figure \ref{sc12}), it is observed that the maximum value of O/N$_2$ ratio had been ($\sim$ 0.5) during May 26-29, 2017. Northern part of the subcontinent showed lower values ($\sim$ 0.3) of O/N$_2$ ratio on May 30 and 31, 2017. Hence no significant change in O/N$_2$ ratio could be observed as compared to the October 2016 storm discussed earlier.

\begin{figure}
\centering\includegraphics[width=1\linewidth]{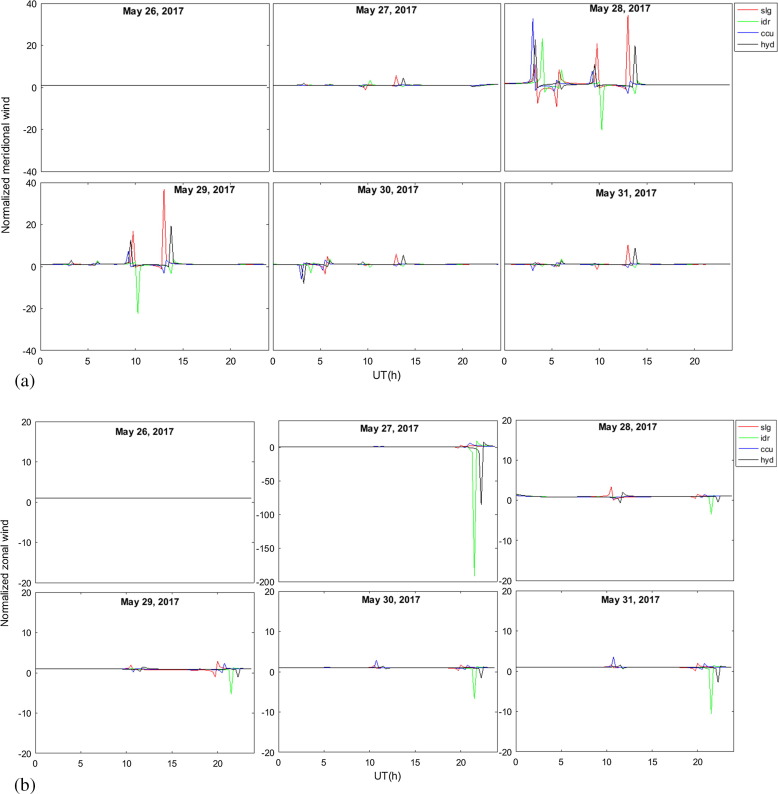}
\caption{Normalized diurnal variations of (a) meridional neutral wind and (b) zonal neutral wind during May 26–31, 2017 for the stations Siliguri (slg in red), Indore (idr in green), Calcutta (ccu in blue) and Hyderabad (hyd in black). (For interpretation of the references to color in this figure legend, the reader is referred to the web version of this article.)}
\label{sc13}
\end{figure}

Figure \ref{sc13}a and b shows the normalized meridional and zonal neutral wind plots, respectively, w.r.t. the quietest day (May 26, 2017) in this period. In Figure \ref{sc13}a, it can be observed that the meridional wind was negative (southward) around 10-11 UT on May 28, 2017 over Indore. This southward wind brings ionized particles towards the lower latitudes of India and hence TEC enhancement is observed over Indore on May 29, 2017. In Figure \ref{sc13}b, on May 27, 2017 around 21-22 UT at Indore and around 22-23 UT at Hyderabad, huge values of negative (westward) zonal wind is observed that signifies that wind is carrying ions westward. Thus it adds additional component to westward equatorial electrojet at nighttime and the net downward E$\times$B drift is strengthened. Hence at Indore and Hyderabad there was no additional enhancement in TEC on May 28, 2017. Thus neutral wind could be another factor for TEC enhancement observed for some of the storm days.

\begin{figure}
\centering\includegraphics[width=1\linewidth]{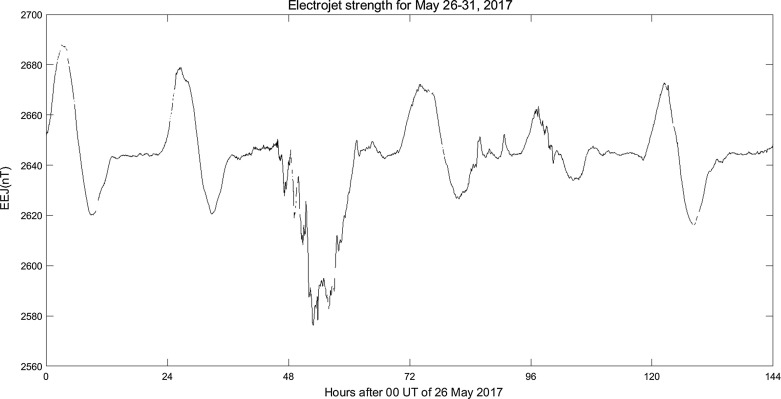}
\caption{Electrojet strength during May 26–31, 2017, taking the difference of horizontal component of the Earth's magnetic field (H in nT) of Alibag (station away from magnetic equator) from that of Dalat (a station near magnetic equator).}
\label{sc14}
\end{figure}

Additionally, it has been observed from Figure \ref{sc14} that on May 28, the difference of H-components between stations Dalat and Alibag decreased to $\sim$2576 nT in comparison to May 27 when the value was $\sim$2620 nT. This decrease in electrojet strength had reduced the E$\times$B force on equatorial ionization, thus reducing distribution of ionization at the off-equatorial regions. This supports the reason behind reduced TEC values as observed in Figure \ref{sc10}.

The TEC response during the main phase of this storm on May 28 and the disturbance on May 29 observed from the Indian longitude shows that there is no significant enhancement during May 27-28. But a rise of TEC in excess of 10-20 TECU was observed around the time of diurnal maximum on May 29.
As AE remained above 600 nT for a duration of 5h on May 28, 2017, TEC enhancement on May 29 may also be explained the AE value in the form of DDEF. The difference in response for the two phases of this storm may be attributed to the differences in their time of onset- first one in the local pre-midnight sector (21:30 LT) and the second one in the local daytime sector (15:30 LT). Since O/N$_2$ ratio did not show any appreciable change during May 26-31 and there were no HSSW event during this period, their contribution to the positive storm effect observed during this period could be eliminated. The importance of the May 2017 storm lies in the fact of its contrasting nature from the October 2016 storms. Although the storm of May 2017 had been more intense (i.e. lower Dst value) in comparison to the October 2016 storms, dynamism of the latter had been greater due to the unique presence of CME and HSSW.  

\section{Conclusions}

In the present paper, dramatic increase in low latitude TEC has been observed in response to the October 12-20, 2016 and May 26-31, 2017 storms from the Indian subcontinent. The October 2016 event presents a unique case of CME-induced storm (October 13-14) followed by a CIR-induced storm (October 16-17) during the declining phase of solar cycle 24. Drastic TEC enhancement of 20-45 TECU in diurnal maximum was observed during the CIR-induced storm of October 2016 event due to Joule heating induced composition change and CIR-induced HSSW event. The May 2017 storm presents a contrasting case of October storms where no composition change or HSSW events have been observed but TEC enhancement of 10-20 TECU in diurnal maximum was observed. Additionally neutral wind values obtained from the HWM14 show positive TEC enhancement from different stations for both the October 2016 and May 2017 storms. Thus even during low solar activity, geomagnetic storm induced electrodynamics at low latitudes is capable of producing TEC enhancements of about 20-45 TECU in excess of quiet time values. This translates to about 3–7m range error at GPS L1 frequency and is detrimental for GNSS applications resulting in degradation of GNSS receiver performance. To the best of our knowledge, such TEC enhancements, from spatially distributed stations in the Indian longitude sector, due to composition change and CIR event during a geomagnetic storm, occurring in the declining phase of the present solar cycle, has not been reported earlier.

\section*{Acknowledgments} 

S. Chakraborty acknowledges the support provided by the Indian Institute of Technology, Indore through research fellowship. The authors also acknowledge Dr. Gopi Krishna Seemala, Indian Institute of Geomagnetism (IIG), Navi Mumbai, India for providing the GPS data analysis program for analysis of the IGS data available at http://sopac.ucsd.edu/dataBrowser.shtml. Acknowledgements go to World Data Center for Geomagnetism, Kyoto for the Dst and AE index. The solar wind and interplanetary magnetic field data are obtained from NASA's SPDF omniweb service https://omniweb.gsfc.nasa.gov
/form/omni\_min.html. The GUVI data used is taken from http://guvitimed.jhuapl
.edu/data\_products while the ground based magnetometer data has been obtained from http://www.intermagnet.org/index-eng.php. S. Ray acknowledges Professor Christine Amory Mazaudier and Professor Sandro Radicella, International Center for Theoretical Physics (ICTP) for helpful discussions. The authors thank both the reviewers for providing valuable comments and suggestions to improve the quality of the manuscript.

\bibliographystyle{model2-names.bst}\biboptions{authoryear}

\bibliography{article.bib}

\end{document}